\documentclass[twocolumn]{webofc}
\usepackage[varg]{txfonts}   
\usepackage{slashed}
\usepackage{subcaption}
\usepackage{array}
\usepackage{makecell}
\usepackage{caption}
\usepackage{amsmath,amssymb}
\usepackage{ulem}
\usepackage{color}

\allowdisplaybreaks

\begin{document}
\title{Hyperon neutrinoless double beta decays in chiral perturbation theory}

\author{
\firstname{Zi-Ying} \lastname{Zhao}\inst{1,2}\thanks{email: zhaoziying@hnu.edu.cn (Speaker)} \and
\firstname{Ze-Rui} \lastname{Liang}\inst{3,1}\thanks{email: liangzr@hebtu.edu.cn}  \and
\firstname{Feng-Kun} \lastname{Guo}\inst{4,5,6}\thanks{email: fkguo@itp.ac.cn} \and
\firstname{Li-Ping} \lastname{He}\inst{7}\thanks{email: heliping@csu.edu.cn} \and
\firstname{De-Liang} \lastname{Yao}\inst{1,2}\thanks{email: yaodeliang@hnu.edu.cn}
}

\institute{
School of Physics and Electronics, Hunan University, Changsha 410082, China
\and
Hunan Provincial Key Laboratory of High-Energy Scale Physics and Applications, Hunan University, Changsha 410082, China
\and
College of Physics and Hebei Key Laboratory of Photophysics Research and Application, Hebei Normal University, Shijiazhuang, Hebei 050024, China
\and
Institute of Theoretical Physics, Chinese Academy of Sciences, Beijing 100190, China
\and
School of Physical Sciences, University of Chinese Academy of Sciences, Beijing 100049, China
\and
Southern Center for Nuclear-Science Theory (SCNT), Institute of Modern Physics, Chinese Academy of Sciences, Huizhou 516000, China
\and
School of Physics, Central South University, Changsha 410083, China
}

\abstract{We study the neutrinoless double beta decays of spin-1/2 hyperons, \(B_1^- \to B_2^+ \ell^- \ell^-\) with $B_1^-\in \{\Sigma^-,\Xi^-\}$ and $B_2^+\in\{p,\Sigma^+\}$, which violate lepton number by two units. The decay amplitudes are computed within covariant baryon chiral perturbation theory at the one-loop level. Dalitz plots of the squared amplitude for six physical decay processes are presented in this proceeding, while a detailed numerical analysis will be presented in a forthcoming publication. 
}
\maketitle
\section{Introduction}
\label{intro}
Neutrinoless double beta $(0\nu\beta\beta)$ decay is one of the most promising experimental processes for determining whether neutrinos are of Majorana or Dirac nature. Observation of this decay would indicate lepton number violation (LNV), shed light on the origin of neutrino masses, and may provide insights into the matter-antimatter asymmetry in the universe. While most studies focus on nuclear $0\nu\beta\beta$ decays~\cite{KamLAND-Zen:2024eml,Majorana:2022udl,CUORE:2019yfd,EXO-200:2019rkq,AMoRE:2024loj,GERDA:2020xhi}, $0\nu\beta\beta$ decays of hyperons provide complementary information and offer an alternative avenue for searching for physics beyond the Standard Model (SM)~\cite{BESIII:2020iwk,BESIII:2025ylz,HyperCP:2005sby}.

In these proceedings, we investigate the $0\nu\beta\beta$ decays of spin-1/2 hyperons, $B_1^-(p_1) \to B_2^+(p_2) \ell^-(k_1) \ell^-(k_2)$, within the framework of covariant baryon chiral perturbation theory (BChPT)~\cite{Gasser:1983yg,Bernard:1995dp}. This approach enables a systematic and model-independent calculation of the decay amplitudes while respecting the underlying symmetries of quantum chromodynamics (QCD) at low energies. We employ the dimension-5 Weinberg operator as the LNV source~\cite{Weinberg1979} and compute the amplitude at the one-loop level. We focus on presenting the decay phase-space distributions using Dalitz plots, which illustrate the variation of the squared amplitude across the kinematically allowed regions. Theoretical predictions of differential decay rates and branching ratios will be provided in a forthcoming work~\cite{zhao:2026xxx}, serving as a benchmark for future searches for LNV hyperon decays.

\section{Theoretical framework}
\subsection{Chiral effective Lagrangian}

The chiral effective Lagrangian for our calculation reads
\begin{align}
\mathcal{L}_{\rm eff} = \mathcal{L}_{M}^{(2)} + \mathcal{L}_{B}^{(1)} + \mathcal{L}_{\Delta L=2}\ ,
\end{align}
where the superscripts denote the chiral orders. In the mesonic sector, the interactions are described by the standard leading-order (LO) chiral Lagrangian~\cite{Gasser:1983yg}:
\begin{align}
\mathcal{L}_{M}^{(2)}=\frac{F_0^2}{4}{\rm Tr}[(D_{\mu}U)^{\dagger}D^{\mu}U]+\frac{F^2_0}{4}{\rm Tr}[U^{\dagger}\chi+U\chi^{\dagger}]\ ,
\label{eq:LagM}
\end{align}
where {$U=\exp(i \phi/F_0)$} with $\phi$ denoting the Goldstone boson (GB) fields. The covariant derivative and $\chi$ are defined as
\begin{align}
D_{\mu}U=\partial_{\mu}U-il_{\mu}U+iUr_{\mu}\ ,	\quad \chi=2B_0(s+ip)\ ,
\end{align}
where $s,p,l_{\mu},r_{\mu}$ represent external scalar, pseudoscalar, left-handed, and right-handed vector sources, respectively. As usual, the light quark mass matrix is incorporated by setting $s={\rm diag}(m_u,m_d,m_s)$ and $p=0$. The constant $B_0$ is related to the quark condensate.

The interactions between GBs and the baryon octet $B$ are described by the chiral Lagrangian~\citep{Krause:1990xc}:
\begin{align}
\mathcal{L}^{(1)}_{B}=\langle \bar{B}\left({\rm i} \slashed{D}-m\right)B\rangle
-\sum_{\pm}\frac{c_{\pm}}{2}\langle \bar{B}\gamma^\mu\gamma_5[u_\mu, B]_\pm\rangle\ ,
\end{align}
where $m$ is the baryon mass in the chiral limit; the anti-commutator and the commutator are accompanied by the low-energy constants (LECs) $c_+=D$ and $c_-=F$. The covariant derivative acting on the baryon field is
\begin{align}
D_{\mu}B=\partial_{\mu}B+[\Gamma_{\mu},B]\ ,
\end{align}
with the chiral connection $\Gamma_{\mu}$ and the vielbein $u_{\mu}$ given by
\begin{align}
\Gamma_{\mu}&=\frac{1}{2}[u^{\dagger}(\partial_{\mu}-ir_{\mu})u+u(\partial_{\mu}-il_{\mu})u^{\dagger}]\ ,\quad u=\sqrt{U},	\\
u_{\mu}&=i[u^{\dagger}(\partial_{\mu}-ir_{\mu})u-u(\partial_{\mu}-il_{\mu})u^{\dagger}]\ .
\end{align}
For the charged-current weak interactions, we identify the left-handed and right-handed sources as
\begin{align}
l_{\mu}&=-2\sqrt{2}G_FT_+[\bar{\nu}_L\gamma_\mu\ell_L]+ {\rm h.c.}, \quad	r_{\mu}=0,
\end{align}
where $T_+$ comprises the elements of the Cabibbo-Kobayashi-Maskawa (CKM) matrix:
\begin{align}
T_+=\left(\begin{array}{ccc}
0  & V_{ud} & V_{us}\\
0  & 0 & 0 \\
0  & 0 & 0
\end{array}\right)\ .
\end{align}
The dimension-5 Weinberg operator in the SM effective field theory gives rise to a Majorana mass term for the left-handed neutrinos~\cite{Weinberg1979}: 
\begin{align}
\mathcal{L}_{\Delta L=2}=-\frac{1}{2}m_{\beta\beta}\nu_{L\beta}^T C\nu_{L\beta}^{}+{\rm h.c.}\ ,
\end{align}
where $m_{\beta\beta}$ is the effective Majorana neutrino mass of $\beta$ type with $\beta\in\{e,\mu,\tau\}$; the superscript $T$ stands for the transpose in the Dirac space.

\subsection{Decay amplitude}

In this subsection, we calculate the one-loop amplitude for $\Delta L = 2$ hyperon decays $B_1^-(p_1) \to B_2^+(p_2)\ell^-(k_1)\ell^-(k_2)$, mediated by light Majorana neutrino exchange. 
It is well known that the $0\nu\beta\beta$ decay at the nuclear level involves the simultaneous conversion of two neutrons into two protons, i.e., $nn\to ppe^-e^-$ as a subprocess of $(A,Z) \to (A,Z+2)+2e^-$. This process is observable when the binding energy of the parent nucleus exceeds that of the intermediate nucleus $(A,Z+1)$, making single beta decay energetically forbidden.
In contrast, at the hadronic level, it is impossible for an isolated neutron to convert into a proton and simultaneously emit two electrons without violating charge conservation. That is, realizing two weak LNV vertices inside an isolated neutron would require two quark-level conversions $d \to u$, transforming $udd$ into $uuu$, a $\Delta^{++}$ configuration. However, the transition $n \to \Delta^{++}+e^-+e^-$ is strictly forbidden kinematically. 

\begin{figure}[htbp]
\centering
\includegraphics[width=0.38\textwidth]{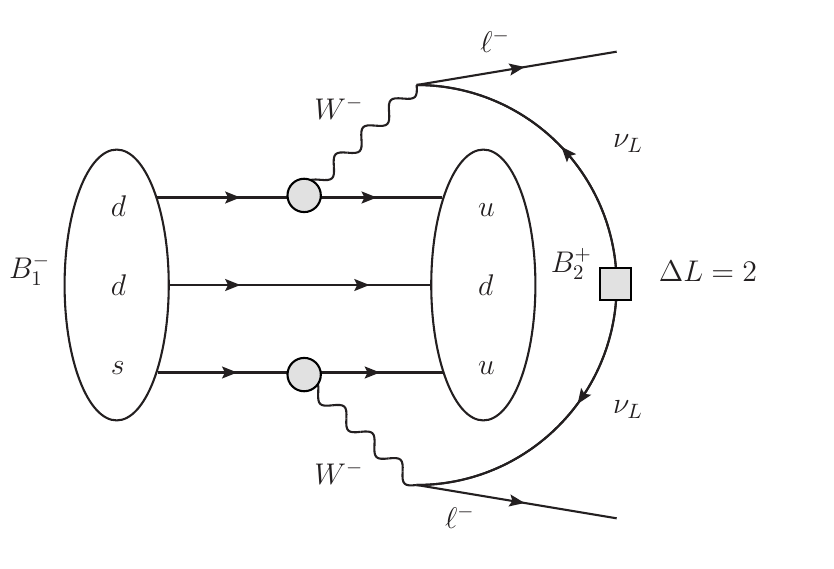}
\caption{Illustration of the hyperon $0\nu\beta\beta$ decay, taking the $\Sigma^- \to p\ell^- \ell^-$ process with $\ell=e,\mu$ as an example.}
\label{fig:quark_level}
\end{figure}
Nevertheless, for hyperons containing a strange quark, the presence of $d$ and $s$ quarks allows the simultaneous conversion of $d \to u$ and $s \to u$ inside the baryon, accompanied by the emission of two charged leptons. Such quark-level flavor transitions can occur without violating charge conservation, provided that the final-state baryon is kinematically accessible. The mechanism of hyperon $0\nu\beta\beta$ decay is illustrated in figure~\ref{fig:quark_level}, where the valence quarks of the initial hyperon undergo two separate charged-current transitions mediated by virtual $W$ bosons. Taking the $\Sigma^-$ hyperon for example, it has the valence quark configuration $dds$. Its LNV decay mechanism is initiated by the simultaneous charged-current transitions of the $s$ quark and one of the $d$ quarks into $u$ quarks. Mediated by the exchange of a virtual Majorana neutrino, these transitions lead to a $\Delta L=2$ process.

\begin{figure}[ht]
\centering
\includegraphics[width=0.9\linewidth]{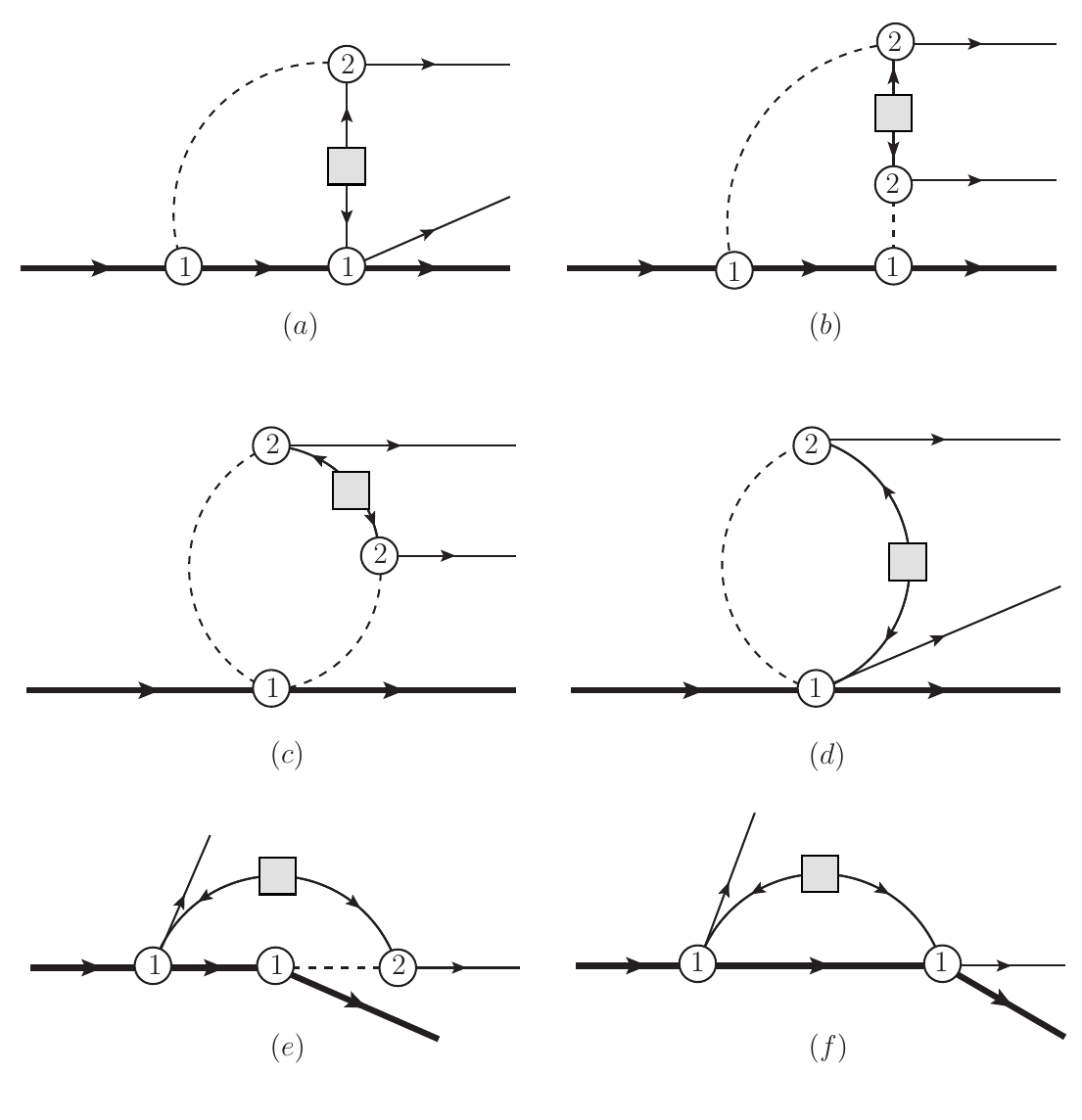}
\caption{One-loop Feynman diagrams contributing to the $0\nu \beta\beta$ decay of hyperons. The thick solid, thin solid and dashed lines represent baryons, leptons and pions. The gray box denotes the $\Delta L= 2$ vertices. The crossed diagrams are not shown explicitly.}
\label{fig:loop}
\end{figure}

The $0\nu\beta\beta$ hyperon decays, allowed by kinematics and the mechanism of figure~\ref{fig:quark_level}, include  
\begin{align}
\Sigma^- &\to p \, e^-e^-,\quad \Sigma^- \to p \, \mu^-\mu^-,  \quad \Sigma^-\to \Sigma^+e^-e^-,  \notag\\
\Xi^- &\to p \, e^-e^-,  \quad   \Xi^- \to p \, \mu^-\mu^-,\quad  \Xi^-\to \Sigma^+e^-e^-.
\label{eq:decay_modes}
\end{align}
Their one-loop Feynman diagrams in BChPT are shown in figure~\ref{fig:loop}. For easy reference, the intermediate states in the loops for each process are specified in table~\ref{tab:intermediate_states}. 
Note that the amplitudes of diagrams (c) and (d) vanish, and we do not show their intermediate states in table~\ref{tab:intermediate_states}.
\begin{table}[ht]
\centering
\caption{Intermediate states inside the loops for the hyperon $0\nu\beta\beta$ decays. For diagrams involving two mesons, we list their symbols from left to right.}
\begin{tabular}{|>{\centering\arraybackslash}p{0.4cm}|>{\centering\arraybackslash}p{1.4cm}|>{\centering\arraybackslash}p{1.4cm}|>{\centering\arraybackslash}p{1.4cm}|>{\centering\arraybackslash}p{1.4cm}|}
\hline
& $\Sigma^- \to p$ & $\Sigma^- \to \Sigma^+$ & $\Xi^- \to p$ & $\Xi^- \to \Sigma^+$ \\
\hline
(a) & \makecell[c]{$n\,K^-$\\$\Sigma^0\,\pi^-$\\ $\Lambda\,\pi^-$} &
\makecell[c]{$\Sigma^0\,\pi^-$\\ $\Lambda\,\pi^-$} &
\makecell[c]{$\Sigma^0\,K^-$\\ $\Lambda\,K^-$} &
\makecell[c]{$\Xi^0\,\pi^-$\\ $\Sigma^0\,K^-$\\ $\Lambda\,K^-$} \\
\hline
(b) & \makecell[c]{$n\,K^-\,\pi^-$\\ $\Sigma^0\,\pi^-\,K^-$\\ $\Lambda\,\pi^-\,K^-$} &
\makecell[c]{$\Sigma^0\,\pi^-\,\pi^-$\\ $\Lambda\,\pi^-\,\pi^-$} &
\makecell[c]{$\Sigma^0\,K^-\,K^-$\\ $\Lambda\,K^-\,K^-$} &
\makecell[c]{$\Xi^0\,\pi^-\,K^-$\\ $\Sigma^0\,K^-\,\pi^-$\\ $\Lambda\,K^-\,\pi^-$} \\
\hline
(e) & \makecell[c]{$n\,\pi^-$\\$\Sigma^0\,K^-$\\$\Lambda\,K^-$} &
\makecell[c]{$\Sigma^0\,\pi^-$\\ $\Lambda\,\pi^-$} &
\makecell[c]{$\Sigma^0\,K^-$\\ $\Lambda\,K^-$} &
\makecell[c]{$\Xi^0\,K^-$\\ $\Sigma^0\,\pi^-$\\ $\Lambda\,\pi^-$} \\
\hline
(f) & \makecell[c]{$n$\\ $\Sigma^0$\\ $\Lambda$} &
\makecell[c]{$\Sigma^0$\\ $\Lambda$} &
\makecell[c]{$\Sigma^0$\\ $\Lambda$} &
\makecell[c]{$\Xi^0$\\ $\Sigma^0$\\ $\Lambda$} \\
\hline
\end{tabular}
\label{tab:intermediate_states}
\end{table}

\begin{table}[htb]
\caption{Explicit expressions of the weak coefficient $T_\text{lept}$.}
\label{tab:Tlept}
\centering
\renewcommand{\arraystretch}{1.2}
\renewcommand{\tabcolsep}{0.8pc}
\begin{tabular}{ccc}
\hline
\multicolumn{1}{c}{$\Delta S=0$} & \multicolumn{1}{c}{$\Delta S=1$} & \multicolumn{1}{c}{$\Delta S=2$} \\
\hline
$\Sigma^- \rightarrow \Sigma^+ e^- e^-$ & $\Sigma^- \rightarrow p\, e^- e^-$ & $\Xi^- \rightarrow p\, e^- e^-$ \\
& $\Sigma^- \rightarrow p\, \mu^- \mu^-$  & $\Xi^- \rightarrow p\, \mu^- \mu^-$   \\
& $\Xi^- \rightarrow \Sigma^+ e^- e^-$ & \\ \hline
$8m_{\beta\beta}G_F^2V^2_{ud}$	&	$8m_{\beta\beta}G_F^2V_{ud}V_{us}$	&	8$m_{\beta\beta}G_F^2V^2_{us}$\\
\hline
\end{tabular}
\end{table}
The decay amplitude $\mathcal{M}$ can be written as 
\begin{align}
\mathcal{M}=&{\rm T}_\text{lept}H_{\mu\nu}L^{\mu\nu}  \ ,\quad L^{\mu\nu}=\bar{u}_{\ell L}(k_1)\gamma^{\mu}\gamma^{\nu} C \bar{u}_{\ell L}^T(k_2)\ ,
\label{eq:M}
\end{align}
where the coefficient ${T}_\text{lept}$ for a given process can be found in table~\ref{tab:Tlept}. In the table, the processes are classified according to the amount of change in strangeness $S$, i.e., $\Delta S$. The hadronic tensor ${H}_{\mu\nu}$ can be decomposed as
\begin{align}
H_{\mu\nu}(s,t,u)=\bar{u}(p_2)\sum_{i=1}^{34}\big[{V}_i\mathcal{O}_{V,\mu\nu}^i+{A}_i\mathcal{O}_{A,\mu\nu}^{i}\big]u(p_1)\ ,
\end{align}
where Mandelstam variables are $s=(k_1+k_2)^2,\ t=(k_1+p_2)^2,\ u=(k_2+p_2)^2$. The Lorentz operators $\mathcal{O}^i_{\mu\nu}$ are
\begin{align}
\label{eq:OV}
\mathcal{O}^1_{V,\mu\nu}&=g_{\mu\nu}, &
\mathcal{O}^2_{V,\mu\nu}&= g_{\mu\nu}\slashed{k}_1, &
\mathcal{O}^3_{V,\mu\nu}&=\gamma_{\mu}\gamma_{\nu}\slashed{k}_1, \notag\\[0mm]
\mathcal{O}^4_{V,\mu\nu}&=\gamma_{\mu}\gamma_{\nu}, &
\mathcal{O}^5_{V,\mu\nu}&=k_{1\mu}k_{2\nu}, &
\mathcal{O}^6_{V,\mu\nu}&=k_{1\nu}k_{2\mu}, \notag\\[0mm]
\mathcal{O}^7_{V,\mu\nu}&=k_{1\mu}k_{1\nu}, &
\mathcal{O}^8_{V,\mu\nu}&=k_{2\mu}k_{2\nu}, &
\mathcal{O}^9_{V,\mu\nu}&=p_{1\mu}p_{1\nu}, \notag\\
\mathcal{O}^{10}_{V,\mu\nu}&=p_{1\mu}k_{1\nu}, &
\mathcal{O}^{11}_{V,\mu\nu}&=p_{1\mu}k_{2\nu}, &
\mathcal{O}^{12}_{V,\mu\nu}&=k_{1\mu}p_{1\nu}, \notag\\
\mathcal{O}^{13}_{V,\mu\nu}&=k_{2\mu}p_{1\nu}, &
\mathcal{O}^{14}_{V,\mu\nu}&=k_{1\mu}k_{2\nu}\slashed{k}_1, &
\mathcal{O}^{15}_{V,\mu\nu}&=k_{2\mu}k_{1\nu}\slashed{k}_1, \notag\\
\mathcal{O}^{16}_{V,\mu\nu}&=k_{1\mu}k_{1\nu}\slashed{k}_1, &
\mathcal{O}^{17}_{V,\mu\nu}&=k_{2\mu}k_{2\nu}\slashed{k}_1, &
\mathcal{O}^{18}_{V,\mu\nu}&=p_{1\mu}p_{1\nu}\slashed{k}_1, \notag\\
\mathcal{O}^{19}_{V,\mu\nu}&=p_{1\mu}k_{1\nu}\slashed{k}_1, &
\mathcal{O}^{20}_{V,\mu\nu}&=p_{1\mu}k_{2\nu}\slashed{k}_1, &
\mathcal{O}^{21}_{V,\mu\nu}&=k_{1\mu}p_{1\nu}\slashed{k}_1,\notag \\
\mathcal{O}^{22}_{V,\mu\nu}&=k_{2\mu}p_{1\nu}\slashed{k}_1, &
\mathcal{O}^{23}_{V,\mu\nu}&=p_{1\mu}\gamma_{\nu}, &
\mathcal{O}^{24}_{V,\mu\nu}&=\gamma_{\mu}p_{1\nu}, \notag\\
\mathcal{O}^{25}_{V,\mu\nu}&=k_{1\mu}\gamma_{\nu}, &
\mathcal{O}^{26}_{V,\mu\nu}&=k_{2\mu}\gamma_{\nu}, &
\mathcal{O}^{27}_{V,\mu\nu}&=\gamma_{\mu}k_{1\nu}, \notag\\
\mathcal{O}^{28}_{V,\mu\nu}&=\gamma_{\mu}k_{2\nu}, &
\mathcal{O}^{29}_{V,\mu\nu}&=p_{1\mu}\gamma_{\nu}\slashed{k}_1, &
\mathcal{O}^{30}_{V,\mu\nu}&=p_{1\nu}\gamma_{\mu}\slashed{k}_1, \notag\\
\mathcal{O}^{31}_{V,\mu\nu}&=k_{1\mu}\gamma_{\nu}\slashed{k}_1, &
\mathcal{O}^{32}_{V,\mu\nu}&=k_{2\nu}\gamma_{\mu}\slashed{k}_1, &
\mathcal{O}^{33}_{V,\mu\nu}&=k_{2\mu}\gamma_{\nu}\slashed{k}_1, \notag\\
\mathcal{O}^{34}_{V,\mu\nu}&=k_{1\nu}\gamma_{\mu}\slashed{k}_1, & &
\end{align}
while $\mathcal{O}_{A,\mu\nu}^i$ can be obtained by $\mathcal{O}_{A,\mu\nu}^i = \mathcal{O}_{V,\mu\nu}^i\gamma_5$ $(i=1,\cdots,34)$. The structure functions $V_i$ and $A_i$ encode the chiral dynamics at one-loop accuracy, calculated from the Feynman diagrams in figure~\ref{fig:loop}. Note that, to remedy the power counting breaking problem~\cite{Gasser:1987rb}, we employ the extended-on-mass-shell (EOMS) scheme~\cite{Fuchs:2003qc}, which has been successfully applied to various processes (see, e.g., Refs.~\cite{Geng:2008mf,Yao:2016vbz,Liang:2023scp}).

It should be emphasized that, since the final-state leptons are identical fermions, the full amplitude must satisfy Fermi--Dirac statistics and be antisymmetric under their  exchange. Consequently, the $u$-channel contribution (corresponding to the diagrams shown in figure~\ref{fig:loop}) acquires a relative minus sign with respect to the $t$-channel amplitude (from the crossed diagrams). In addition, with the help of the identity
\begin{align}
\label{eq:su}
\bar{u}_{\ell L}(k_2)\gamma^{\nu}\gamma^{\mu} C\bar{u}_{\ell L}^T(k_1)=-\bar{u}_{\ell L}(k_1)\gamma^{\mu}\gamma^{\nu} C \bar{u}_{\ell L}^T(k_2)\ ,
\end{align}
the full hadronic amplitude can be established as 
\begin{align}
H_{\mu\nu}=H_{\mu\nu}^u+H_{\mu\nu}^t\ ,\quad  H_{\mu\nu}^t =  H_{\nu\mu}^u (k_1\leftrightarrow k_2, u\leftrightarrow t)\ .
\end{align}

\section{Results}
\label{sec:Results}

Using Casimir's tracing technique, the amplitude squared ${|\mathcal{M}|^2}$ can be deduced to 
\begin{align}
{|\mathcal{M}|^2}&=|{\rm T}_\text{lept}|^2{\rm Tr}[\gamma^{\mu}\gamma^{\nu}\slashed{k}_2 P_L \gamma^{\beta}\gamma^{\alpha}\slashed{k}_1 P_R] \notag\\
&\times {\rm Tr}  [H_{\mu\nu}(\slashed{p}_1+m_{1})\overline{H_{\alpha\beta}}(\slashed{p}_2+m_{2})],
\end{align}
where $P_R=\frac{1+\gamma^5}{2},P_L=\frac{1-\gamma^5}{2}$ and $\overline{H_{\alpha\beta}}=\gamma^0 H_{\alpha\beta}^{\dagger} \gamma^0$.

In our numerical computation, the CKM matrix elements are taken to be $|V_{ud}|=0.97367$, and $|V_{us}|=0.22431$. We use the Fermi constant $G_F=1.16636 \times 10^{-5}~\rm GeV^{-2}$~\cite{ParticleDataGroup:2024}, the LECs $D=0.8$ and $F=0.5$~\cite{Borasoy:1998pe}. 

\begin{figure*}
\centering
\begin{subfigure}[b]{0.32\textwidth} 
\centering
\includegraphics[width=\textwidth]{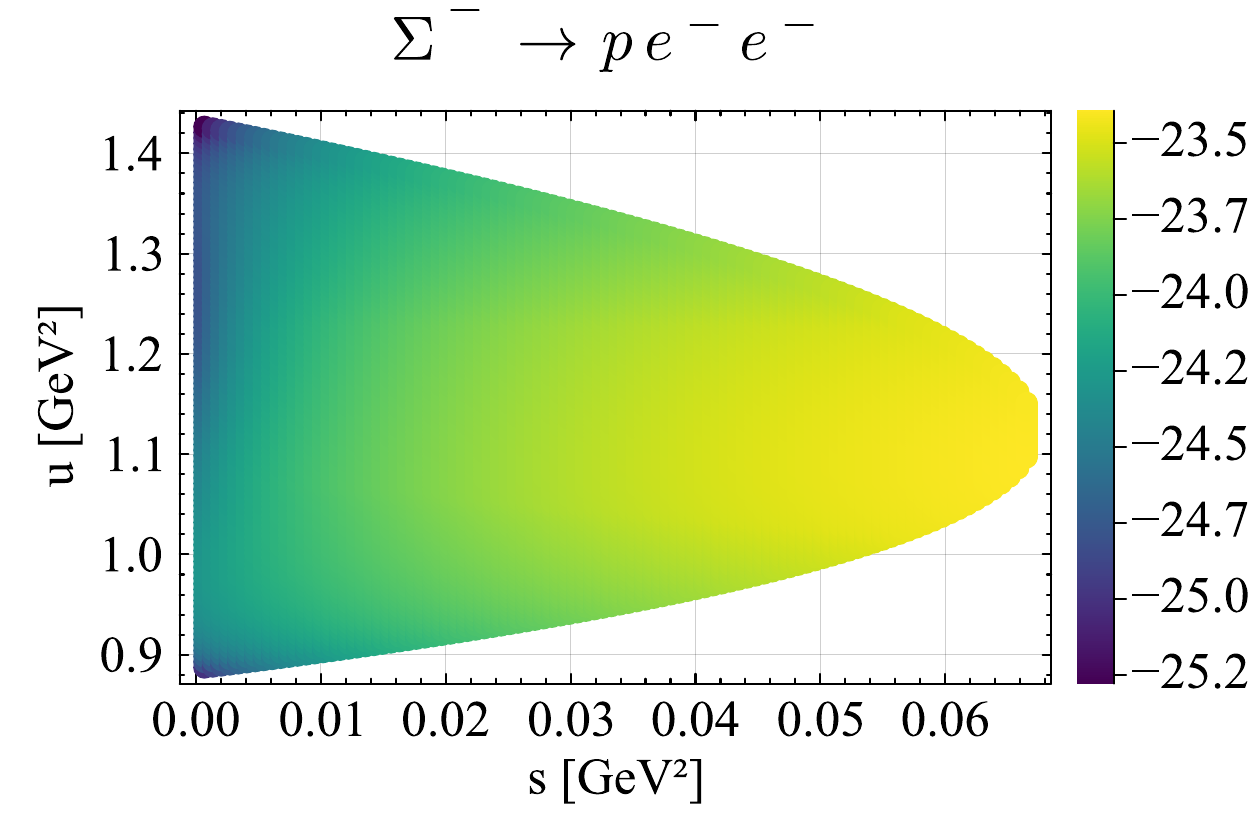}
\label{fig}
\end{subfigure}
~~
\begin{subfigure}[b]{0.32\textwidth}
\centering
\includegraphics[width=\textwidth]{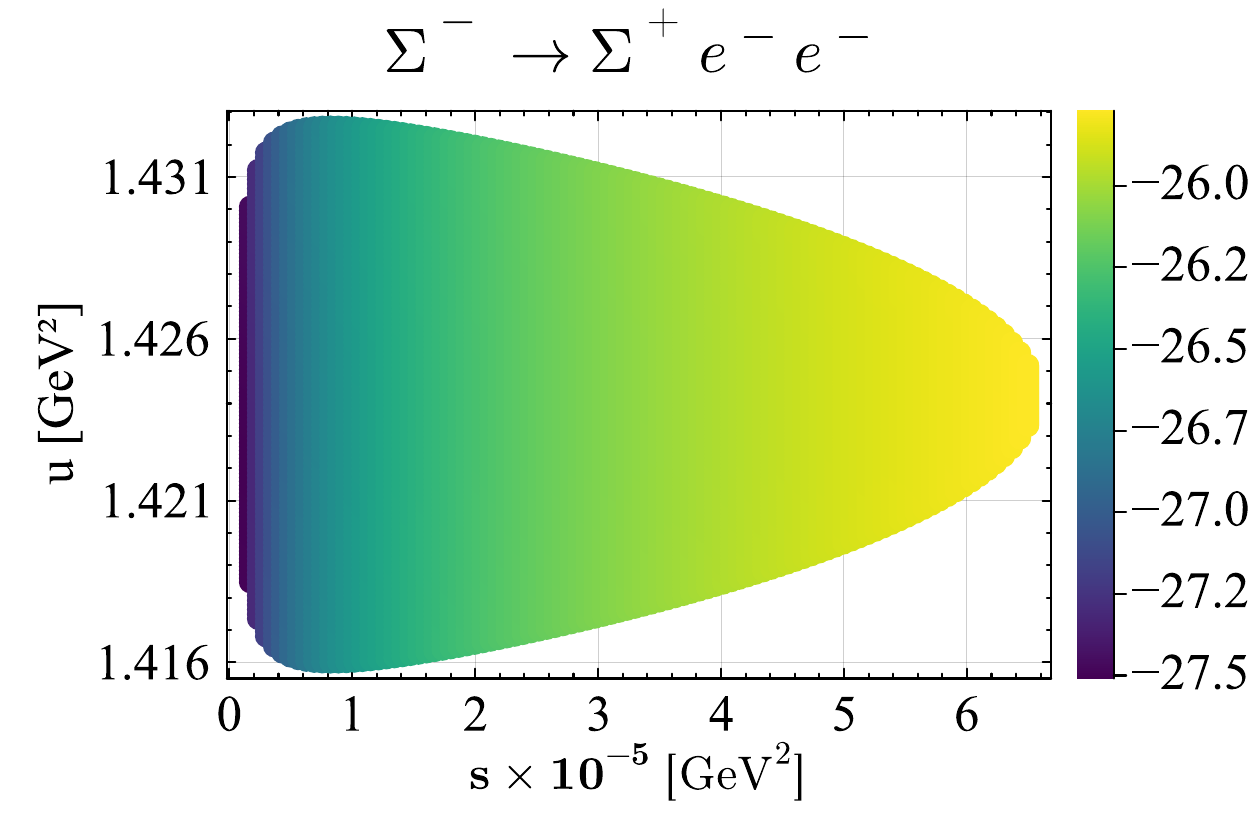}
\label{fig:SSll}
\end{subfigure}
~~
\begin{subfigure}[b]{0.32\textwidth}
\centering
\includegraphics[width=\textwidth]{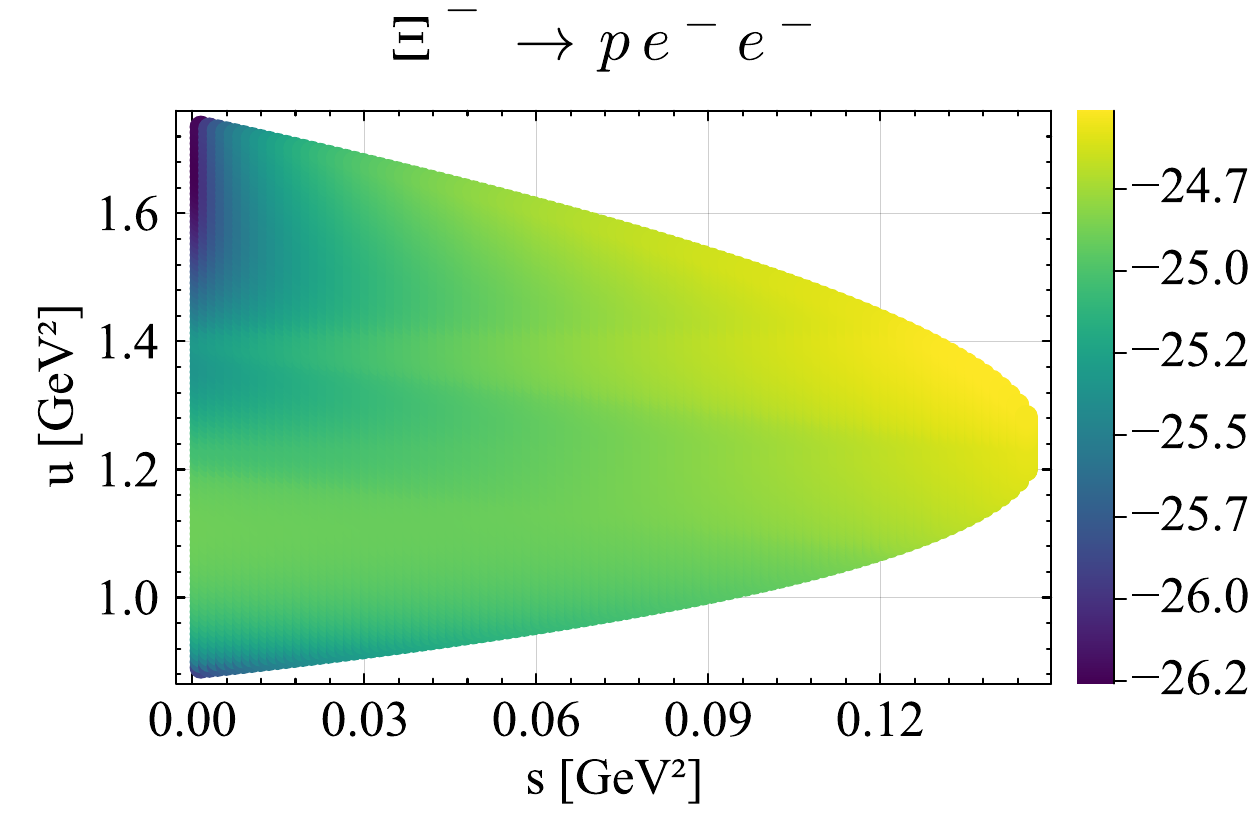}
    \label{fig:Xpll}
\end{subfigure}\\
\begin{subfigure}[b]{0.32\textwidth}
    \centering
    \includegraphics[width=\textwidth]{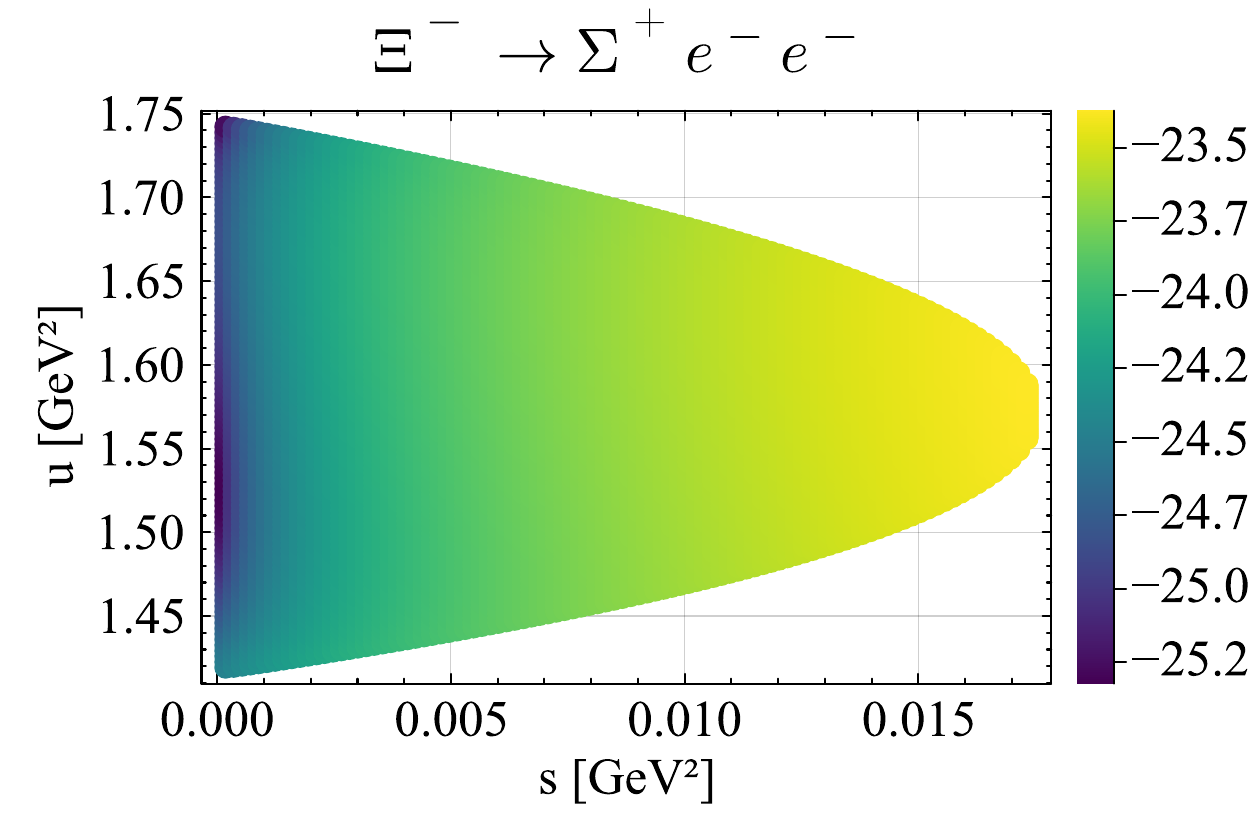}
    \label{fig:XSll}
\end{subfigure}
~~
\begin{subfigure}[b]{0.32\textwidth}
    \centering
    \includegraphics[width=\textwidth]{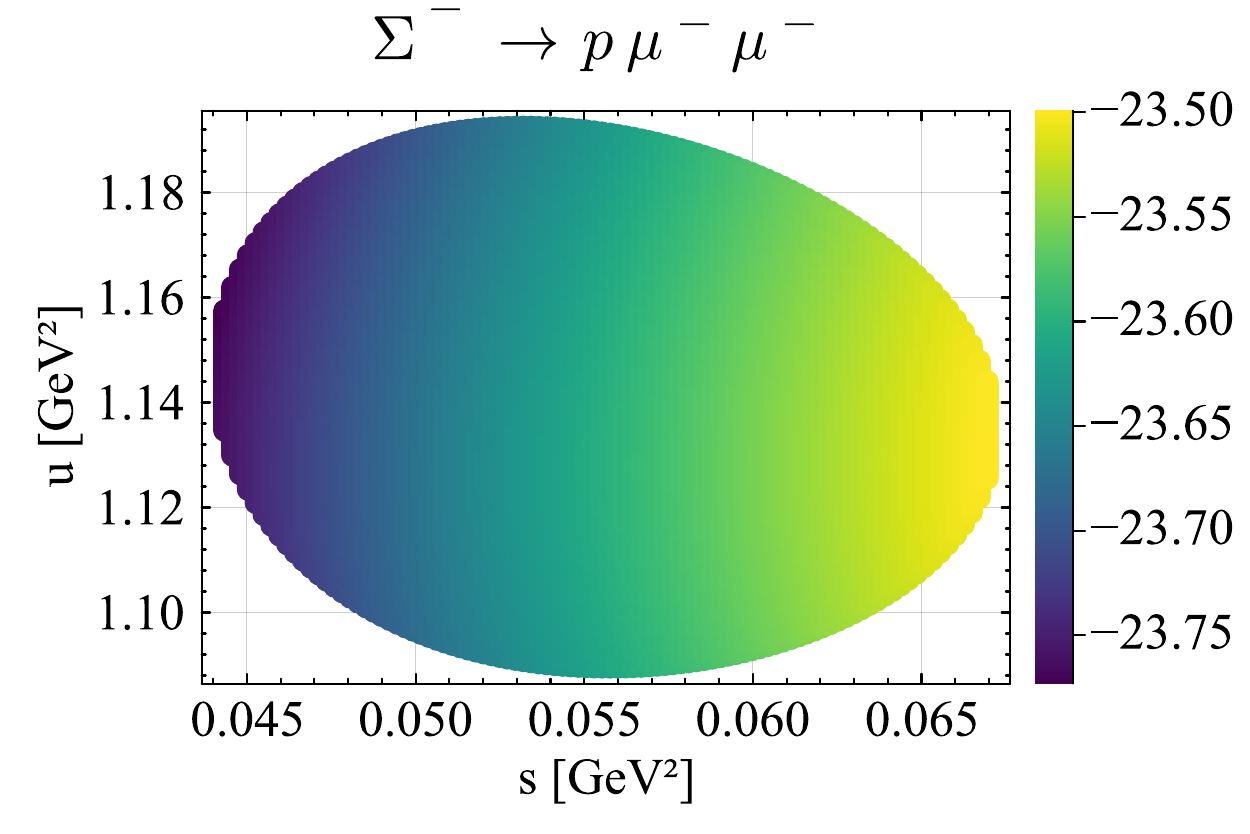}
    \label{fig:Spmu}
\end{subfigure}
~~
\begin{subfigure}[b]{0.32\textwidth}
    \centering
    \includegraphics[width=\textwidth]{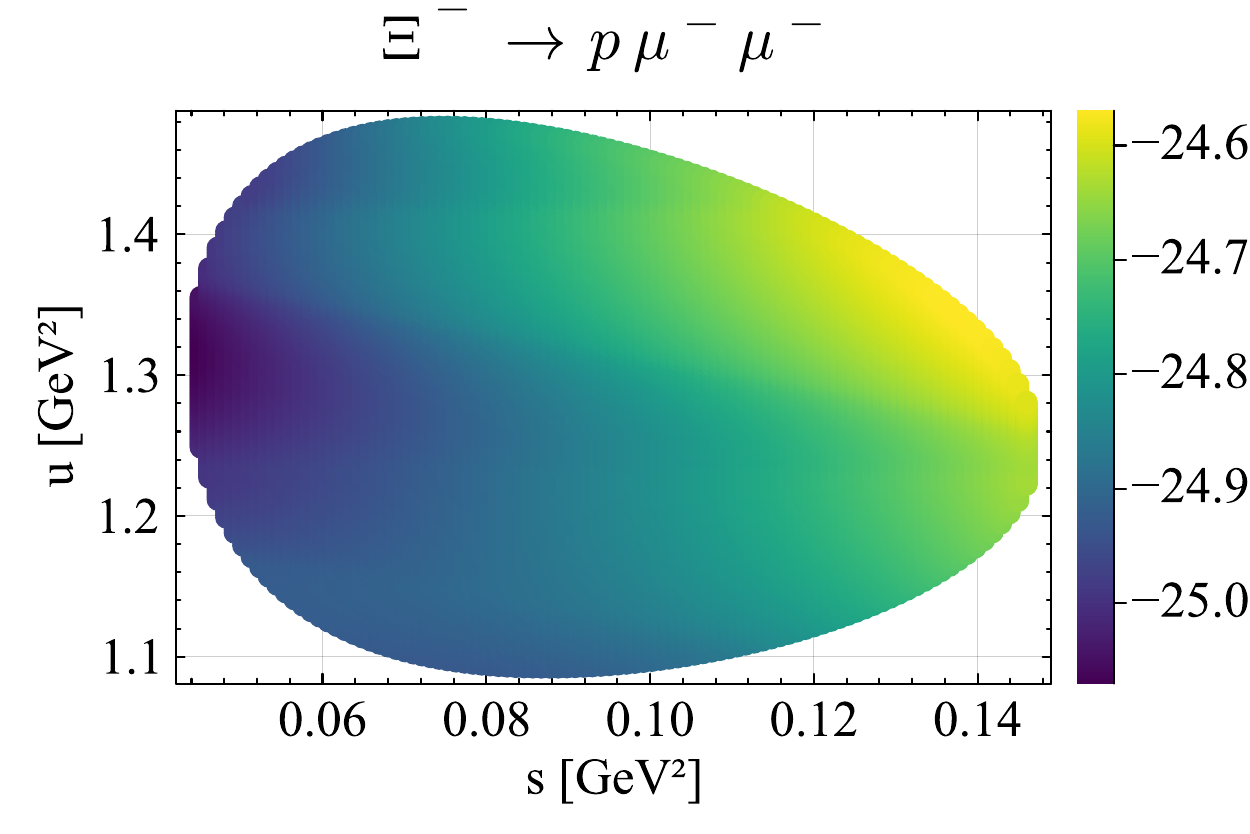}
    \label{fig:Xpmu}
\end{subfigure} 
\caption{Dalitz plots for the 6 hyperon $0\nu\beta\beta$ decay processes, illustrating the variation of the amplitude distribution across the kinematic region of $s$ and $u$. The color scale on the right corresponds to the logarithm of the amplitude modulus squared, i.e., $\log({|\mathcal{M}|^2}/{\langle m_{\beta\beta} \rangle^2})$.}
\label{fig:dalitz}
\end{figure*}

Figure~\ref{fig:dalitz} presents Dalitz plots for the 6 hyperon $0\nu\beta\beta$ decay processes given in Eq.~\eqref{eq:decay_modes}. These plots show the distribution of the amplitude squared over the allowed kinematic phase space, providing a clear visualization of how the amplitude modulus varies with respect to the kinematic variables $s$ and $u$. The differences between the electronic and muonic modes arise from their distinct phase-space and mass effects.

\section{Summary}
\label{sec:Summary}
In summary, with the inclusion of the dimension-5 Weinberg operator, we have performed a systematic calculation of $0\nu\beta\beta$ decays of spin-1/2 hyperons in BChPT at the one-loop level. We derive all the relevant one-loop amplitudes and plot the Dalitz distributions for each decay channel. More results on integrated observables, such as differential decay rates and branching ratios, will be presented in a forthcoming work~\cite{zhao:2026xxx}. Our study provides theoretical results useful for future searches of LNV signatures beyond SM and for exploring the origin of neutrino mass in experiments.

\section*{Acknowledgments}
ZYZ would like to thank the organizers of INPC 2025 for hospitality. This work is supported by National Nature Science Foundations of China under Grants No.~12547166, No.~12275076, No.~12335002, No.~12125507, and No.~12447101; by the Science Fund for Distinguished Young Scholars of Hunan Province under Grant No.~2024JJ2007; by the Science Foundation of Hebei Normal University under Grants No. L2025B09; by Science Research Project of Hebei Education Department under Grant No. QN2025063; by Hebei Natural Science Foundation under Grant No. A2025205018; and by CAS under Grant No. YSBR-101.


\bibliography{biblio}

\end{document}